# Hydrogenated Bilayer Wurtzite SiC Nanofilms: A Two-Dimensional Bipolar Magnetic Semiconductor Material

**Long Yuan, Zhenyu Li, Jinlong Yang***

Hefei National Laboratory for Physical Sciences at Microscale, University of Science and Technology of China, Hefei, 230026, China

Recently, a new kind of spintronics materials, bipolar magnetic semiconductor (BMS), has been proposed. The spin polarization of BMS can be conveniently controlled by a gate voltage, which makes it very attractive in device engineering. Now, the main challenge is finding more BMS materials. In this article, we propose that hydrogenated wurtzite SiC nanofilm is a two-dimensional BMS material. Its BMS character is very robust under the effect of strain, substrate, or even a strong electric field. The proposed two-dimensional BMS material paves the way to use this promising new material in an integrated circuit.

## 1. Introduction

Semiconductor-based spintronics recently have attracted lots of attentions, which seeks to utilize the spin degree of freedom of the carriers and leads to a revolution in the current electronic information processing technology.[1] There are two big challenges in spintronics. One is generating spin polarized carriers, and the other is manipulating it. The first challenge can be met if we have a kind of materials, where its one spin channel is metallic while the other is semiconducting. This kind of materials is called half metal, and various half-metallic nanostructures have been proposed in previous study.[2-8]

With totally spin polarized carriers provided by half metallic materials, to manipulate the spin polarization direction, typically an external magnetic field is required. However, we notice that an electric control of spin is more attractive, since electric field can be easily applied locally. Recently, such a goal has been realized in a newly proposed spintronics material, bipolar magnetic semiconductor (BMS).[9] In BMS, the valence and conduction bands possess opposite spin polarization when approaching the Fermi level. Therefore, completely spin-polarized currents with reversible spin polarization can be created and controlled simply by applying a gate voltage. A prototype one dimensional (1D) BMS material has been proposed previously.[9] The remain open question is if the BMS character can widely exist, for example, in two dimensional (2D) materials. For applications in integrated circuits, a two-dimensional (2D) BMS material is more desirable.

Here, we report a study on hydrogenated wurtzite SiC nanofilm, which is predicted to be a 2D BMS material. There are several merits with this new BMS material. First of all, it is transition metal (TM) free. Traitionally, magnetism in semiconductors is typically introduced by doping magnetic impurities,[10] such as TMs. However, considering the spin-scattering problem in the electron transmission, TM-free materials are more promising. On the other hand, compared with the extensively studied 2D material, graphene or silicon,[11-17] SiC is attractive for spintronics applications due to the larger spin-orbit interaction of Si than C.[18-20] Thirdly, the material proposed in this study is expected to be readily synthesized in experiment. Actually, graphitic SiC sheet has been formed by sonication of wurtzite SiC,[21] and our first-principles molecular dynamics (MD) simulations suggest that the graphitic-like SiC bilayer could be easily transformed back into the wurtzite structure via surface hydrogenation. Therefore, we provide a feasible way to control the spin polarization of carriers in two-dimensional materials by electric means.

## 2. Computational methods

All geometry optimizations and electronic structures calculations were performed by spin-polarized density functional theory (DFT) as implemented in the Vienna *ab initio* Simulation Package (VASP). The project-augmented wave method for core-valence interaction and the generalized gradient approximation (GGA) of the Perdew-Burke-Erzerhof (PBE) form for the exchange-correlation function was employed.[22-25] A kinetic energy cutoff of 500 eV was chosen in the plane-wave expansion. A large value (15 Å) of the vacuum region was used to avoid the interaction between two adjacent periodic images. Reciprocal space was represented by Monkhorst-Pack special k-point scheme.[26] The Brillouin zone was sampled by a set of 11×11×1 k-points for the geometry optimization and 21×21×1 k-points for the static total

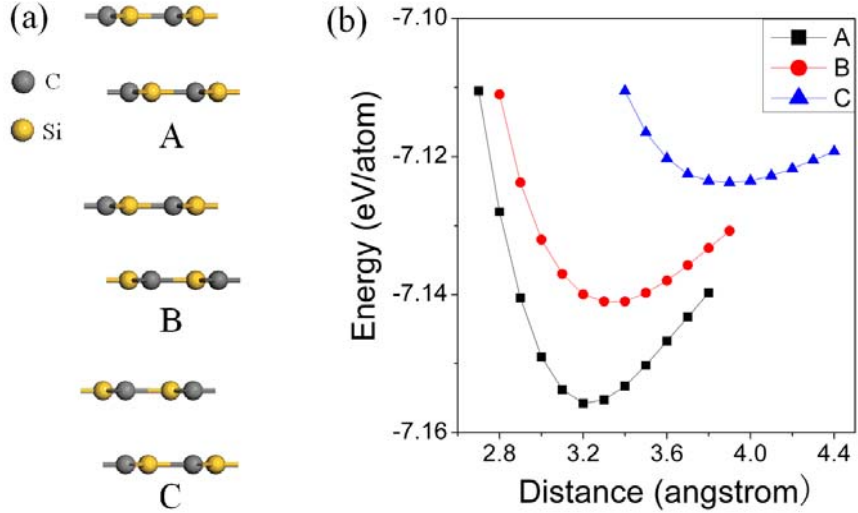

Fig 1. (a) Graphitic-like SiC bilayers with three types of stacking orderings (A, B and C), (b) Total energy for the A, B, C stacking arrangements as a function of distance between the two layers, in eV/atom.

energy calculations. To investigate the magnetic coupling of the hydrogenated nanofilms, we performed calculations using a 2×2 supercell. The Monkhorst-Pack special k-point of 6×6×1 and 11×11×1 were employed for the geometry optimization and static total energy calculations. The structure was relaxed using conjugate gradient scheme without any symmetry constraints and the convergence of energy and force was set to $1\times10^{-4}$ eV and 0.01 eV/Å. The accuracy of our methods was tested by calculating the C-Si bond length and band gap of pristine SiC sheet. Our calculated results are 1.789 Å and 2.54 eV, in good agreement with previous theoretical results.[19, 27]

It is well known that GGA usually fails to accurately describe the van der Waals (vdW) interactions. So, the empirical correction method of Grimme (DFT+D2) was employed for this part of interaction.[28, 29] Standard parameters of the dispersion coefficients $C_6$ (0.14, 1.75, and 9.230 J nm$^6$ mol$^{-1}$, for H, C, and Si, respectively), vdW radii (1.001, 1.452, and 1.716 Å), cutoff radius (30.0 Å), global scaling factor (0.75 Å), and damping factor d (20.0 Å) have been used. Local density approximation (LDA) was also used to compute bilayer distance and agreed well with the PBE (with vdW correction) results.

The external electric field, as implemented in VASP code, was introduced by adding an planar dipole layer in the middle of vacuum part in the periodic supercell.[30] Since GGA usually underestimates band gap, we also performed test calculations with the hybrid Heyd-Scuseria-Ernzerhof (HSE) functional.[31, 32] Although band gap increased with HSE functional, the main results reported here remained unchanged.

MD simulations, as implemented in the VASP code, were employed to study the transformation process and thermal stability of the hydrogenated wurtzite SiC nanofilms. A kinetic energy cutoff of 400 eV and PBE functional (with vdW correction) were chosen. MD simulations were performed in NVT ensemble with a 4×4 supercell. 2×2×1 k-points were used to sample the 2D Brillouin zone. MD simulations at 200K (or 450K) lasted for 2 ps (or 10 ps) with a time step of 1.0 fs (or 2.0 fs) were performed to investigate the transformation process (or thermal stability). The temperature was controlled by using the Nose-Hoover method.[33] The climbing image nudged elastic band (CI-NEB) method was used for transition state search.[34, 35] Four or five images were inserted between the initial and final states. Images were optimized until the energy and force on each atom were less than $1\times10^{-4}$ eV and 0.02 eV/Å.

## 3. Results and discussion

In SiC sheets, hydrogenation can be completed by absorbing H atoms on C site or Si site. A 2×2 supercell consisting of four unit cells in the SiC monolayer were constructed to investigate the energetically favorable absorbing site. For this purpose, we defined the formation energies of semihydrogenated SiC monolayer as

$E_f = (E_{H-SiC} - E_{SiC} - m\mu_H)/N$

where $E_{H-SiC}$, $E_{SiC}$ are the energies of the semihydrogenated SiC sheet and pristine SiC sheet, respectively. We choose $\mu_H$ as the energy of a H atom. m, N are the numbers of hydrogen atoms and total number of atoms in the supercell. The optimized bond length of C-H and Si-H are 1.118 and 1.515 Å, respectively. The formation energies of the two semihydrogenated SiC sheets, labeled as H-CSi and H-SiC, are -0.565 and -0.440 eV/atom, respectively. Our results are consistent with previous DFT computations.[36] The formation energy of H-CSi is smaller than H-SiC, indicating that H atom prefers to absorb on the C site.

Next, we examine three types of stacking arrangements in the graphitic-like SiC multilayered structures: (A) C-Si ordering, (B)

**Table 1** Calculated energies and distances for the three types of stacking orderings (A, B and C) by using PBE(with vdW correction) and LDA functional, respectively, in eV/atom, Å.

|   | PBE (with vdW correction) | | LDA | |
|---|---|---|---|---|
|   | Energy | Bilayer distance | Energy | Bilayer distance |
| A | -7.157 | 3.221 | -7.714 | 3.205 |
| B | -7.142 | 3.396 | -7.704 | 3.407 |
| C | -7.124 | 3.889 | -7.687 | 3.981 |

Si-Si ordering, and (C) C-C ordering, as shown in Figure 1a. Based on our calculations, C-Si ordering is the most energetically favorable stacking arrangement, which also agrees with previous theoretical computations.[37] The interlayer spacing value is 3.221 Å. LDA calculations (Table 1) also leads to similar results, with an interlayer spacing of 3.205 Å, slightly smaller than the PBE (with vdW correction) result.

MD simulations are performed to study hydrogenation on the top layer at C sites of the bilayer SiC. Four snapshots in the 2 ps trajectory at 200 K are shown in Figure 2a. In the initial state, the SiC sheets remain planar structures. At the intermediate stage (I), strong σ-bond between C atom and H atom is formed and the original planar $sp^2$-hybridized SiC structure is broken, leaving unpaired electrons on Si atoms. The hydrogenated top layer exhibits a buckling structure and the bottom layer remains intact. Then, at the intermediate stage (II), due to interlayer interaction, the bottom layer gradually exhibits the buckling structure. The distance between the two layers becomes shorter. At last, as shown in the final stage, the interlayer C-Si bond is formed and the transformation is completed. The formed hydrogenated wurtzite nanofilms can be very stable during a subsequent 10 ps MD simulation at 450 K.

We name the hydrogenated wurtzite SiC bilayer as H-$(CSi)_2$. The optimized structure at 0 K is shown in Fig 2b. C atoms and Si atoms in the first layer become fully $sp^3$-hybridized. In the second layer, only the C atoms are $sp^3$-hybridized, the Si atoms remain $sp^2$-hybridized. The optimized lattice parameter is 3.102 Å, slight larger than the pristine SiC sheet. The length of C-H bond and interlayer C-Si bond are 1.108 Å and 1.899 Å, respectively. The minimum energy path (MEP) of the hydrogenation is shown in Fig 2d. The system only need to overcome a small barrier of 0.03 eV/atom to complete the transformation and the formed wurtzite nanofilm is more energetically favorable by about 0.27 eV/atom. Therefore, the hydrogenated wurtzite nanofilm should be feasible to be synthesized in experiment under mild conditions.

To further investigate the structural stability of the hydrogenated nanofilms, another formation energy is defined as

$E_f = (E_{nanofilm} - E_{SiC} - m\mu_H)/n$

where $E_{nanofilm}$, $E_{SiC}$ are the energies of the hydrogenated nanofilms, pristine SiC multilayers, respectively. We choose $\mu_H$ as the energy of a H atom. m, n are the numbers of H atoms and total atoms in the supercell, respectively. The computed formation energy of H-(CSi)$_2$ nanofilm is -0.592 eV/atom. A negative value indicates an exothermic transformation process, providing more evidence for the thermodynamics stability.

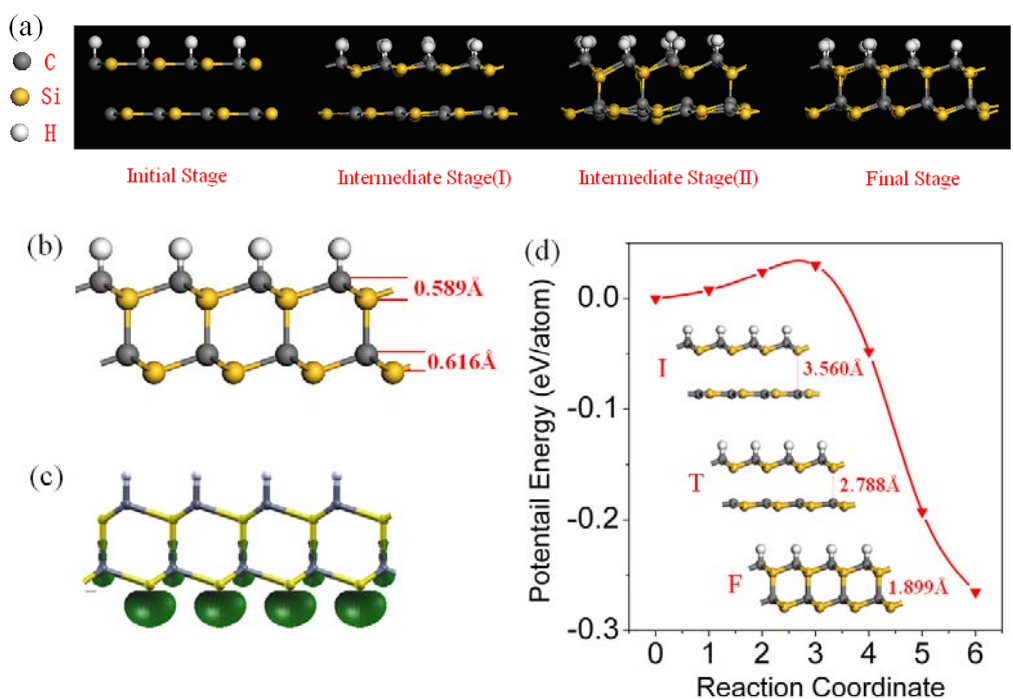

**Fig 2.** Formation of the hydrogenated wurtzite nanofilm from the semihydrogenated SiC bilayer. (a) Snapshots of the initial, intermediate, and final stage of the transformation. (b) Optimized geometry of H-$(CSi)_2$ at 0 K. (c) Spin density distribution of H-$(CSi)_2$ (isosurface value: 0.05 e/Å$^3$). (d) Calculated MEP of the transformation, insets show the atomic structures of initial (I), transition (T), and final (F) states along the energy path.

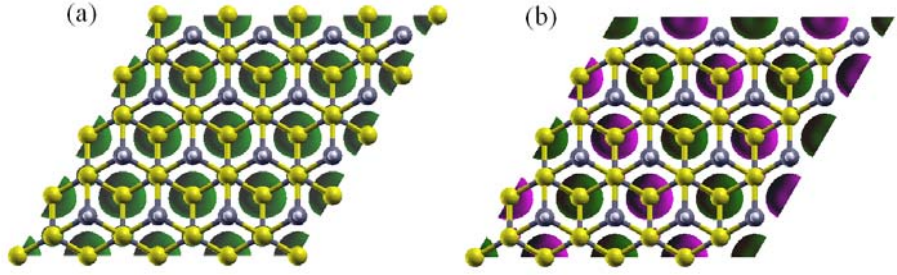

**Fig 3.** (a) Ferromagnetic and (b) antiferromagnetic configurations. Green and pink are used to indicate the positive and negative signs of the spin density (isosurface value: 0.05 e/Å$^3$), respectively.

As we mentioned, the Si atoms in the second layer are still remaining sp$^2$-hybridized, leaving the electrons in the unhydrogenated Si atoms localized and unpaired, which gives about 1 μ$_B$ magnetic moment per unit cell, as shown in Fig 2c. In order to identify the preferred magnetic coupling of these moments, three kinds of magnetic configurations were considered: (1) ferromagnetic (FM) coupling; (2) antiferromagnetic (AFM) coupling; (3) nonmagnetic (NM) coupling. The first two cases are as shown in Figure 3. The calculated results show that the FM state is energetically more favorable than the AFM and NM states. FM coupling is 21 and 273 meV per unit cell lower in energy than AFM and NM coupling.

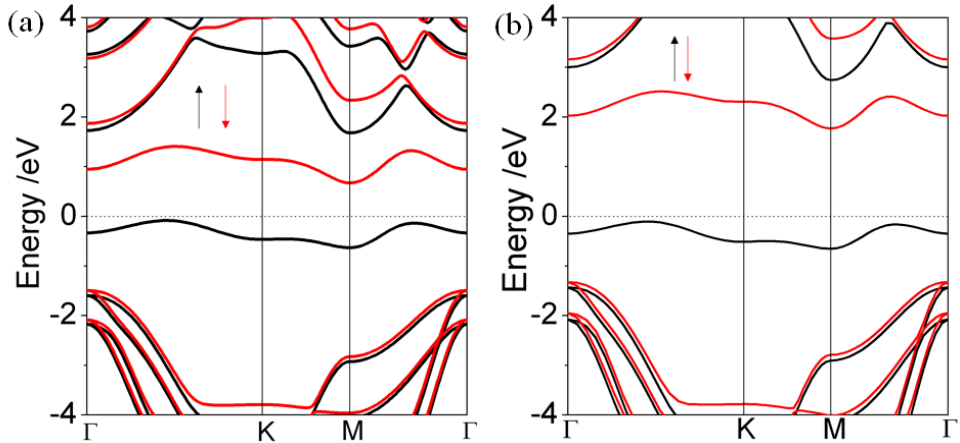

**Fig 4.** Electronic band structures of H-(CSi)$_2$ based on (a) PBE functional and (b) HSE functional. The Fermi level is set at zero.

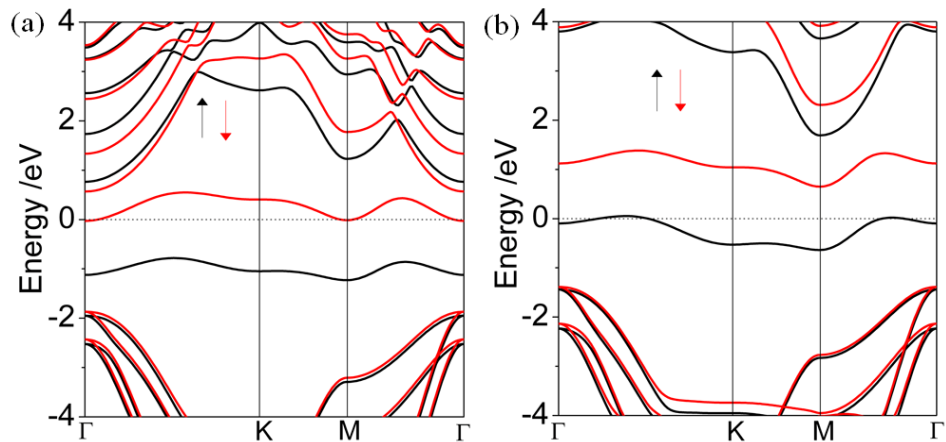

**Fig 5.** Band structures of H-(CSi)$_2$ with doping concentrations of (a) 0.02 electron per atom and (b) 0.03 hole per atom. The Fermi level is set at zero.

As shown in Figure 4a, H-(CSi)$_2$ is a ferromagnetic semiconductor with a small band gap of 0.834 eV. Our test calculation with HSE functional also gives similar band structure (Figure 4b), with an enlarged band gap of 1.874 eV. Importantly, the spin polarizations of H-(CSi)$_2$ are different for the valence band (VB) and the conduction band (CB). Therefore, it is a typical BMS material. Half-metallicity with opposite spin polarization can be obtained via electron or hole doping, as shown in Figure 5. Take the doping level of 0.03 hole per atom for example, the Fermi level moves down, crossing the VB. The spin-up channel becomes gapless, while the spin-down channel is still insulating, with a band gap about 2 eV. The FM coupling is still energetically favorable under electron or hole doping. For example, the system with 0.03 hole doping per atom favors the FM state by 16 meV per unit compared to the AFM state. The net magnetic moment becomes 0.85 $\mu_B$ per unit cell, indicating that the doped hole is mainly distributed on the unhydrogenated Si atoms.

An electronic device may under an electric field. So, we first check the effect of external field to H-(CSi)$_2$. The electric field was added perpendicularly to the nanofilm, with a strength ranging from -0.8 to 0.8 V/angstrom. The positive direction of the electric field is defined along the z axis as shown in Figure 6. Under an electric field as strong as 0.4 V/angstrom, the band gap is only slightly affected by the field. Therefore, the H-(CSi)$_2$ is a robust BMS material even under strong electric field. We also observed that the band gap of the spin-up channel decrease rapidly when the electric field is larger than 0.2 V/angstrom. This is an effect of the nearly free electron (NFE) state,[38] which in marked in blue in Figure 6. The NFE state exists widely in the low dimensional materials,[39-42] and it can be easily shifted with electrostatic potential. As a byproduct, an occupied NFE state is realized in our system, which can be used as ideal transport channels, when the electric field is larger than 0.6 v/angstrom.

On the other hand, the hydrogenated wurtzite nanofilm should be on a suitable substrate in applications. A substrate can apply a stress on the material. Then, it is interesting to see if the BMS electronic structure is robust under external strain. To answer this question, we applied external stress to H-(CSi)$_2$, which is defined as $\varepsilon = (c_1 - c_0)/c_0$, where $c_1$ and $c_0$ are the unit cell parameters with and without deformation. The tensile or compression strain is uniformly applied along both in-plane lattice vectors. As shown in Figure 7, the BMS characters in H-(CSi)$_2$ are well survived under external strains of ±3%. Besides, we have also checked that

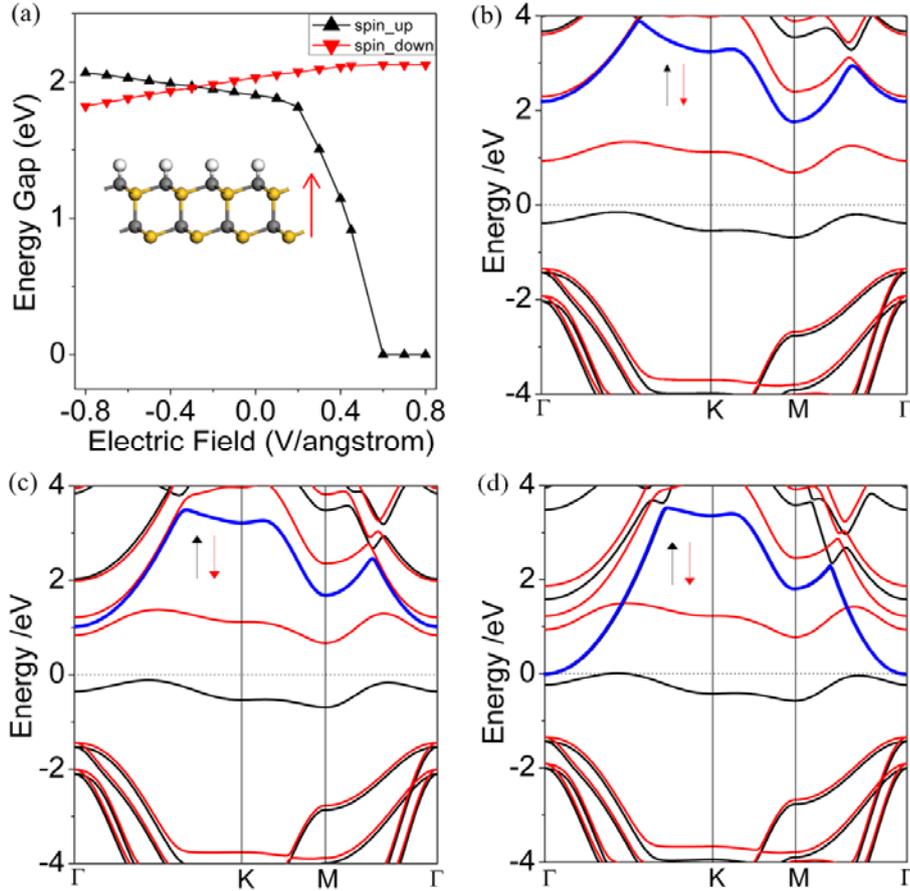

**Fig 6.** (a) Electric field induced energy gap modulation of H-(CSi)$_2$, with the positive direction denoted by the red arrow. Band structures of H-(CSi)$_2$ under an electric field of (b) 0.0 V/angstrom, (c) 0.4 V/angstrom, and (d) 0.6 V/angstrom. The states in blue line are the nearly free electron states. The

Fermi level is set at zero.

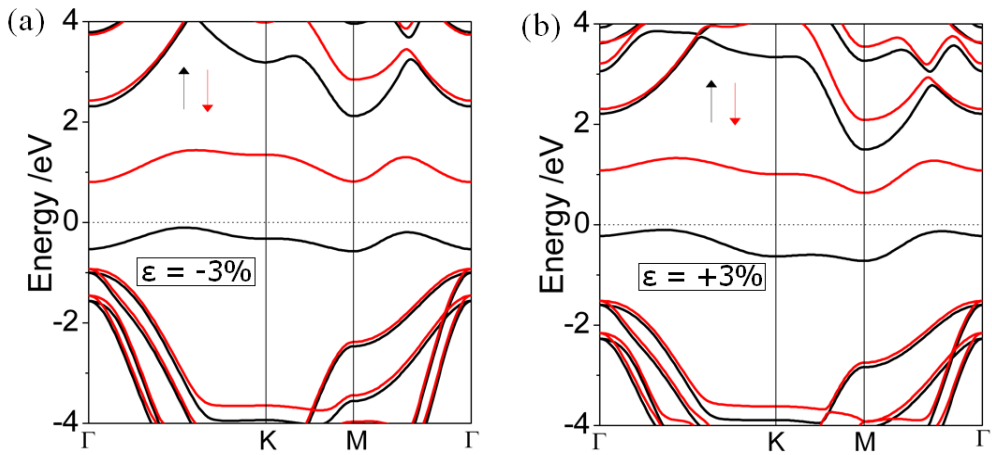

**Fig 7.** Electronic band structures of H-(CSi)$_2$ under external strain. The Fermi level is set at zero.

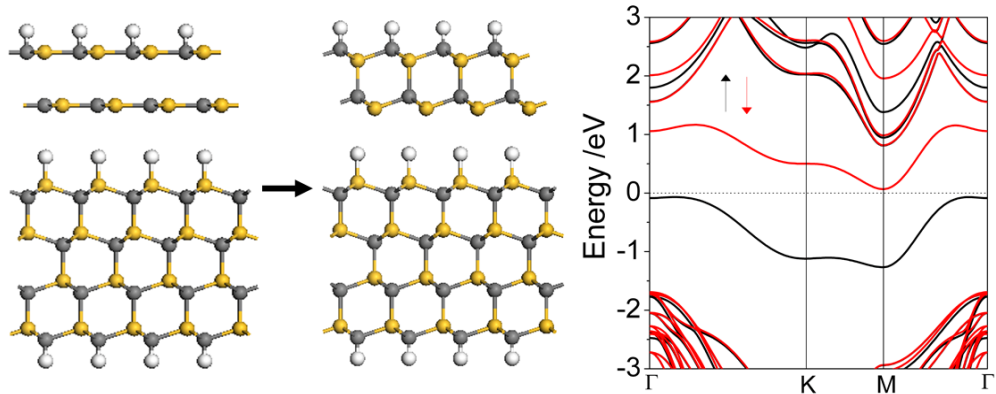

**Fig 8.** The initial and optimized geometry, the band structures of H-(CSi)$_2$ nanofilm on a hydrogenated (0001) SiC substrate. The Fermi level is set at zero.

the FM coupling is still the energetically favorable magnetic configuration under these two stress levels. For example, the energy of FM state is 12 meV per unit cell lower than that of the AFM state at a stress level of +3%. The energy difference becomes smaller compared to H-(CSi)$_2$ without strain. This is reasonable, since the distance between two neighboring magnetic Si atoms is increased under a tensile stress, and the exchange coupling is thus efficiently weakened.

We have also tested the system of semihydrogenated SiC bilayer on a real substrate, the hydrogenated (0001) SiC surface. After DFT+D2 geometry optimization, the distance between the SiC bilayer and substrate is 3.01 Å. For this system, we still get 1 $\mu_B$ magnetic moment per unit cell. More importantly, the BMS character of the electronic structure is still kept, as shown in Figure 8. On the substrate, the band gap between the two spin channel becomes very small (0.136 eV), which makes the gate voltage based spin polarization control very feasible.

## 4. Conclusion

In summary, based on first-principles calculations, we have proposed that hydrogenated SiC bilayer will keep the wurtzite structure and it is a 2D TM-free BMS material, which could introduce half-metallicity with opposite spin polarization via electron or hole doping possibly realized by a simple gate voltage control. This BMS material is robust under moderate external electric field and under relatively high level of external strain possibly induced by a substrate. Our results thus pave the way to spintronics application of the promising new BMS material.

### Acknowledgements.

This work is partially supported by the National Key Basic Research Program (2011CB921404), by NSFC (21121003, 91021004, 20933006), and by USTCSCC, SCCAS, Tianjin, and Shanghai Supercomputer Centers.